\documentclass[superscriptaddress,showpacs,amsmath,amssymb,aps,prd,twocolumn,floatfix,nofootinbib]{revtex4-1}

\usepackage{graphicx}
\usepackage{dcolumn}
\usepackage{bm}
\usepackage{color}
\usepackage{amsmath}
\usepackage{amssymb}
\usepackage{bbm}
\usepackage{lineno}
\usepackage{hyperref}
\usepackage{hhline}
\hypersetup{
    colorlinks = true
}

\usepackage{natbib}
\usepackage{graphicx}
\usepackage{ amssymb }

\usepackage[colorinlistoftodos]{todonotes}
\begin{document}

\title{Point Proposal Network for Reconstructing 3D Particle Endpoints with Sub-Pixel Precision in Liquid Argon Time Projection Chambers}

\newcommand{\SLAC}{SLAC National Accelerator Laboratory, Menlo Park, CA, 94025, USA}
\affiliation{\SLAC}
\newcommand{\ICME}{ICME, Stanford University, Stanford, CA, 94305, USA}
\affiliation{\ICME}
\newcommand{\Stanford}{Stanford University, Stanford, CA, 94305, USA}
\affiliation{\Stanford}

\author{Laura~Domin\'e} \email{ldomine@stanford.edu} \affiliation{\Stanford}
\author{Pierre~C\^ote~de~Soux} \affiliation{\ICME}
\author{Fran\c cois~Drielsma} \affiliation{\SLAC}
\author{Dae~Heun~Koh} \affiliation{\Stanford}
\author{Ran~Itay} \affiliation{\SLAC}
\author{Qing~Lin} \affiliation{\SLAC}
\author{Kazuhiro~Terao} \affiliation{\SLAC}
\author{Ka~Vang~Tsang} \affiliation{\SLAC}
\author{Tracy~L.~Usher} \affiliation{\SLAC}

\collaboration{on behalf of the DeepLearnPhysics Collaboration}\noaffiliation

\begin{abstract}
    Liquid Argon Time Projection Chambers (LArTPC) are particle imaging detectors recording 2D or 3D images of trajectories of charged particles. Identifying points of interest in these images, namely the initial and terminal points of track-like particle trajectories such as muons and protons, and the initial points of electromagnetic shower-like particle trajectories such as electrons and gamma rays, is a crucial step of identifying and analyzing these particles and impacts the inference of physics signals such as neutrino interaction.
    The Point Proposal Network is designed to discover these specific points of interest. The algorithm predicts with a sub-voxel precision their spatial location, and also determines the category of the identified points of interest.
    Using as a benchmark the PILArNet public LArTPC data sample in which the voxel resolution is 3mm/voxel, our algorithm successfully predicted 96.8~\% and 97.8~\% of 3D points within a distance of 3 and 10~voxels from the provided true point locations respectively. For the predicted 3D points within 3 voxels of the closest true point locations, the median distance is found to be 0.25 voxels, achieving the sub-voxel level precision. 
    In addition, we report our analysis of the mistakes where our algorithm prediction differs from the provided true point positions by more than 10~voxels. Among 50 mistakes visually scanned, 25 were due to the definition of true position location, 15 were legitimate mistakes where a physicist cannot visually disagree with the algorithm's prediction, and 10 were genuine mistakes that we wish to improve in the future. 
    Further, using these predicted points, we demonstrate a simple algorithm to cluster 3D voxels into individual track-like particle trajectories with a clustering efficiency, purity, and Adjusted Rand Index of 96~\%, 93~\%, and 91~\% respectively.
\end{abstract}


\maketitle

\section{Introduction}
Accelerator based neutrino oscillation experiments have successfully deployed deep convolutional neural networks (CNN) in their data analysis pipeline~\cite{UBPaper1, UBPaper2, UBNature}. Many of the present and future experiments utilize a liquid argon time projection chamber (LArTPC), a class of particle imaging detectors which produces 2D or 3D images over many meters of detected charged particle trajectories, with a resolution of the order of ~mm/pixel.
Examples of such experiments along with their respective active volumes include MicroBooNE (90 tons)~\cite{MicroBooNE}, the Short Baseline Near Detector (SBND, 112 tons)~\cite{SBN}, ICARUS (600 tons)~\cite{ICARUS} and the Deep Underground Neutrino Experiment (DUNE, 40,000 tons)~\cite{DUNE}.

The particle trajectories recorded in LArTPC images often appear as 1D lines in a 2D or 3D space. Their topological features can be diverse, ranging from straight line-like tracks to branching tree-like electromagnetic showers. In the process of analyzing an image from the pixel-level energy deposits to build a larger picture of particle trajectories with their respective kinematic properties, detecting points of interest such as the initial and terminal points of particle trajectories in the early stage of a data reconstruction chain is critical.
For example, in clustering tasks on electromagnetic (EM) showers, the initial point can help to define a general direction for the whole shower that includes dozens to hundreds of EM secondaries. This is especially useful for separating neutral pions, a source of major background to $\nu_e$ signal for neutrino oscillation analysis as well as an important sample for detector energy calibration, from cosmic rays and neutrino-nucleus interactions. Finding these points can also be a crucial step in determining a neutrino interaction vertex. If each particle trajectory is associated with these points of interests, the predicted points naturally include candidates for the neutrino interaction vertex.

Localizing an arbitrary number of such points in an image is analogous to a task called object detection in the field of Computer Vision. Many object detection algorithms based on CNNs have been proposed~\cite{he2017mask,girshick2015fast,ren2015faster,ssd} including Faster Region Convolutional Neural Network (Faster R-CNN) which has been one of the most popular choices for object detection applications and also successfully applied in LArTPC image data~\cite{UBPaper1}. Faster R-CNN consists of a feature extractor CNN and an attention mechanism called Region Proposal Network (RPN). The feature extractor consists of convolution layers and pooling layers, and generates a data tensor with low spatial resolution compared to the input. The RPN takes this data tensor and generates region proposals, typically rectangular shaped bounding boxes, that are likely to contain a target object in the original image resolution. The insight of RPN is to act on a spatially contracted data tensor which contains fewer pixels compared to the original input, thus addressing the challenge of long compute time. R-CNN is a family of algorithms that employ the RPN concept. One of the most recent of these is Mask R-CNN,~\cite{he2017mask} which is undeniably the most popular object instance detection algorithm to date.


Inspired by the concept behind RPN, we have designed a Point Proposal Network (PPN) to identify points of interest in a LArTPC image, namely the initial point of EM particles, referred to as {\it shower-like} particles in this paper, as well as the initial and terminal points, collectively referred to as {\it endpoints}, of {\it track-like} particles, which include all but shower-like particles. 
While RPN is responsible for predicting both the location and size of a bounding box for an object detection, PPN is simplified to propose only the location as the target is a point, not an object. Our goal is to integrate PPN into a generic, full 3D data reconstruction chain for LArTPCs, which consists of multi-task deep neural networks, such that the whole chain can be optimized end-to-end. Building on the previous effort, we use U-ResNet~\cite{domine2019scalable} as the feature extractor and implement PPN to predict the position and {\it semantic type} of an arbitrary number of points in an image with voxel-level precision. 

The contributions of this paper are two-fold:
\begin{itemize}
    \item Introduce PPN for reconstructing the 3D endpoints of track-like particles and the initial point of shower-like particle with sub-voxel precision.
    \item Provide a performance benchmark on a public LArTPC simulation dataset (PILArNet) for future reference and comparison against other methods.
\end{itemize}
While, in this paper, our target is 3D LArTPC images, the concept of PPN is applicable to both 2D and 3D images~\cite{nu2018poster}. Section \ref{sec:architecture} introduces the architecture of the UResNet network that we use as the backbone for PPN, as well as the details on PPN architecture, the loss definitions and post-processing methods. Section~\ref{sec:experiments} outlines our experiments setup, including details on the public LArTPC data sample that we use. Section~\ref{sec:results} shows a first benchmark of the PPN performance on this sample.

The study presented in this paper is fully reproducible using a \textsc{Singularity}~\cite{Singularity} software container \footnote{\href{https://singularity-hub.org/containers/11757}{https://singularity-hub.org/containers/11757}}, implementations available in the \texttt{lartpc\_mlreco3d}\footnote{\href{https://github.com/DeepLearnPhysics/lartpc_mlreco3d}{https://github.com/DeepLearnPhysics/lartpc\_mlreco3d}} repository and public simulation samples~\cite{opendata} made available by the DeepLearnPhysics collaboration.

\begin{figure*}[t]
    \centering
    \includegraphics[width=0.98\textwidth]{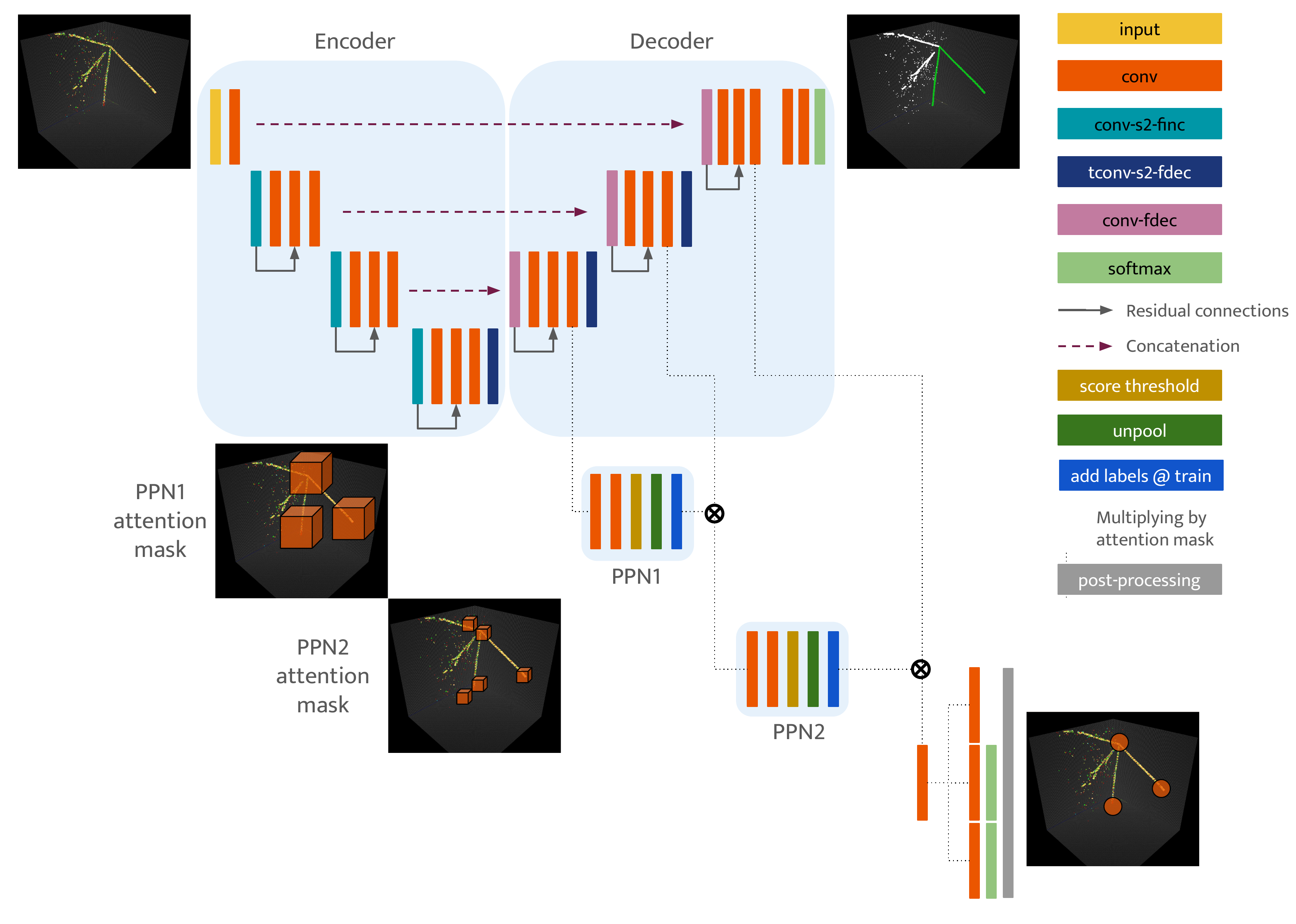}
    \caption{U-ResNet architecture for semantic segmentation. In this example we say that the U-ResNet has a depth of 3 since we perform 3 downsamplings. Turquoise boxes represent convolutions with stride 2 and increasing the number of filters. Dark blue boxes are transpose convolutions with stride 2 and decreasing the number of filters. Purple boxes are convolutions with stride 1 that decrease the number of filters. Blue boxes represent the addition of true voxels to the mask of positive voxels, and only apply during training. The gold boxes are a score thresholding ($>0.5$) operation on the softmax of predicted scores.  The spatial size of the data tensor is constant across the horizontal dimension.}
    \label{fig:U-ResNet}
\end{figure*}

\section{Network Architecture \label{sec:architecture}}
The network architecture consists of two parts: U-ResNet~\cite{domine2019scalable} and PPN. Both blocks include many CNN layers. In order to make our algorithm scalable to a large-scale LArTPC detector analysis, we designed the whole chain using Sparse Submanifold Convolutional Network (SSCN)~\cite{submfdcnn,3dsubmfdcnn}.

\subsection{UResNet: feature extractor}
U-ResNet is designed for a voxel-level classification task, called semantic segmentation, for 3D LArTPC images~\cite{domine2019scalable}. The architecture of U-ResNet can be divided into two parts, namely {\it encoder} and {\it decoder}. The encoder consists of repeated blocks of convolution and strided convolution layers which down-samples the image resolution while increasing the features dimension, thus learning from key features in an image at different scales. We refer to the number of down-sampling operations as \textit{depth}. The decoder takes the low-spatial size, highly compressed data tensor from the encoder and up-samples them back to the original image resolution. After each up-sampling operation, the data tensor of matching spatial size is taken from the encoder output and concatenated to the up-sampled tensor before the combined tensor is further processed by convolution layers in the decoder. The key concept behind the concatenation operation, introduced by the original U-Net~\cite{UNet} authors, is to recover the lost spatial resolution information in the encoder block due to strided convolution layers and effectively combine with the abstract features contained in the up-sampled tensor. Convolution layers in the decoder block are trained to best combine high spatial resolution information and abstract feature information. As a result, they learn how to best spatially interpolate abstract features extracted by the encoder back to the original image resolution. Figure~\ref{fig:U-ResNet} shows the architecture of U-ResNet. For the study carried out in this paper, we set the depth of UResNet to 6 and used 16 filters at the first convolution layer. The number of filters increases linearly with the depth, and is 96 at the deepest layer.

\subsection{PPN layers}
Within the U-shaped network architecture (see Figure~\ref{fig:U-ResNet}), we implement PPN by introducing additional convolution layers at different spatial resolutions, starting with the most contracted data tensor at the lowest spatial resolution. While these PPN layers could be attached to either the encoder or the decoder of U-ResNet, it is more powerful to attach them to the decoder block as data tensors generated by the decoder should be more information rich.
At the deepest level and coarsest spatial resolution, the so-called PPN1 produces a softmax score of a value between 0 and 1 for each voxel, which indicates whether or not the voxel contains the location coordinate of any of the {\it true points}, i.e. the target 3D points to be detected. We call this {\it detection score} in the following.
We call the voxels {\it positive} if the detection score is above the set threshold value. We call other voxels {\it negative}. Positive voxels yield an attention mask that we can use at the next step.
At an intermediate level and medium spatial resolution, we up-sample the mask predicted by PPN1 and use it as an attention mask to pre-select candidate positive voxels. The so-called PPN2 layer then similarly predicts a subset of positive voxels among these pre-selected voxels in the attention mask at this spatial resolution. 
Finally, we up-sample the result of PPN2 to the original image resolution and use it as another attention mask. The final layer, so-called PPN3, is made of 3x3 convolutions which predict the following quantities for each voxel that has been selected through these successive attention masks:
\begin{itemize}
    \item a detection score (of being a voxel within some neighborhood of a ground truth point, for which we choose a distance threshold of 5~voxels),
    \item a 3D position (offset with respect to the voxel center), 
    \item  and a type score (for the point to belong to a semantic type).
\end{itemize}
We note that the 3D position prediction made from a particular voxel may be located in its neighbor voxel. This implies that multiple neighbor voxels of a target voxel, which contains one or more of true points, can all propose positions that are within the target voxel, which gives more information to identify 3D points and can improve the performance.


\begin{figure*}[t]
    \centering
    \includegraphics[width=0.48\textwidth]{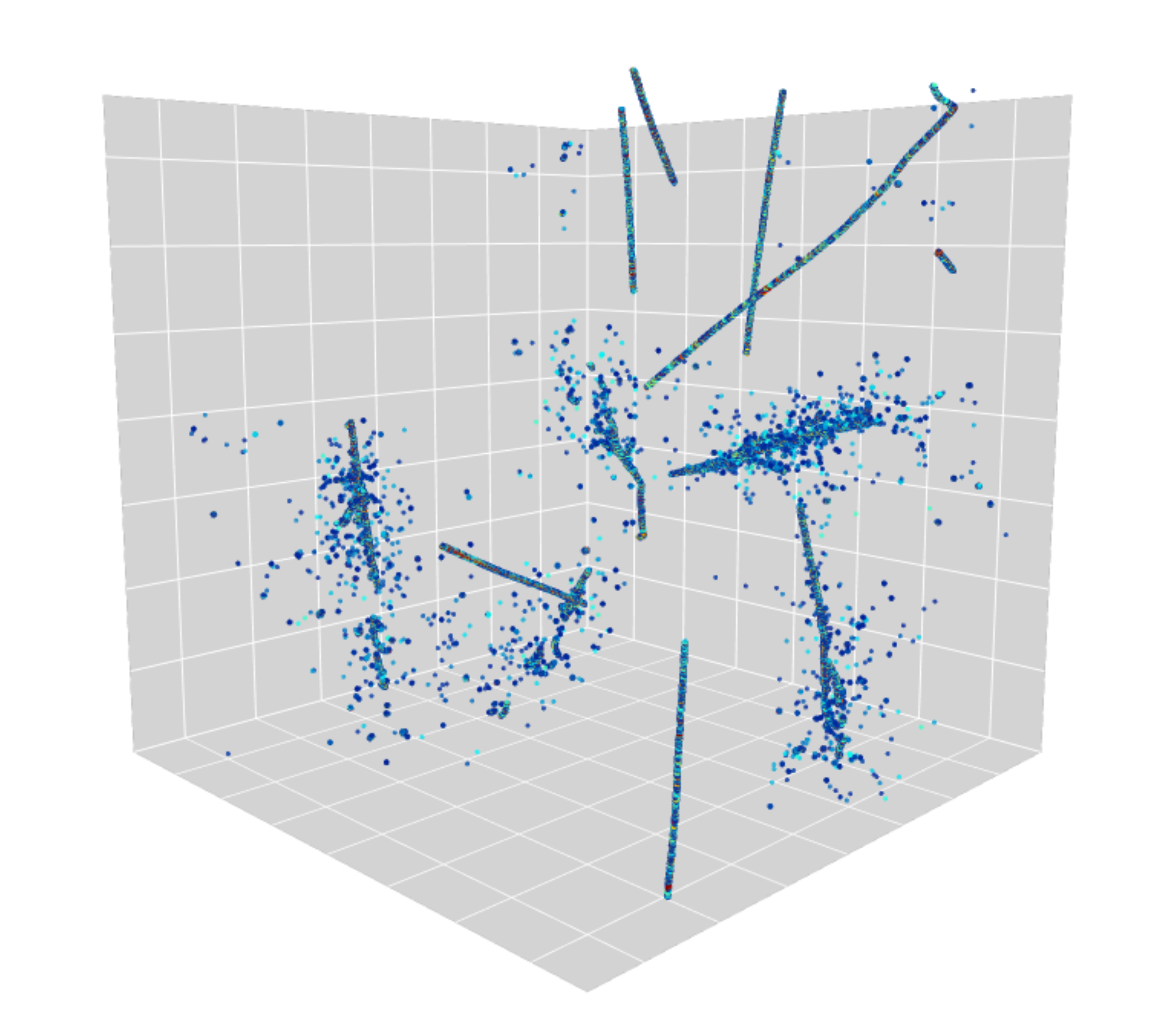}
    \includegraphics[width=0.48\textwidth]{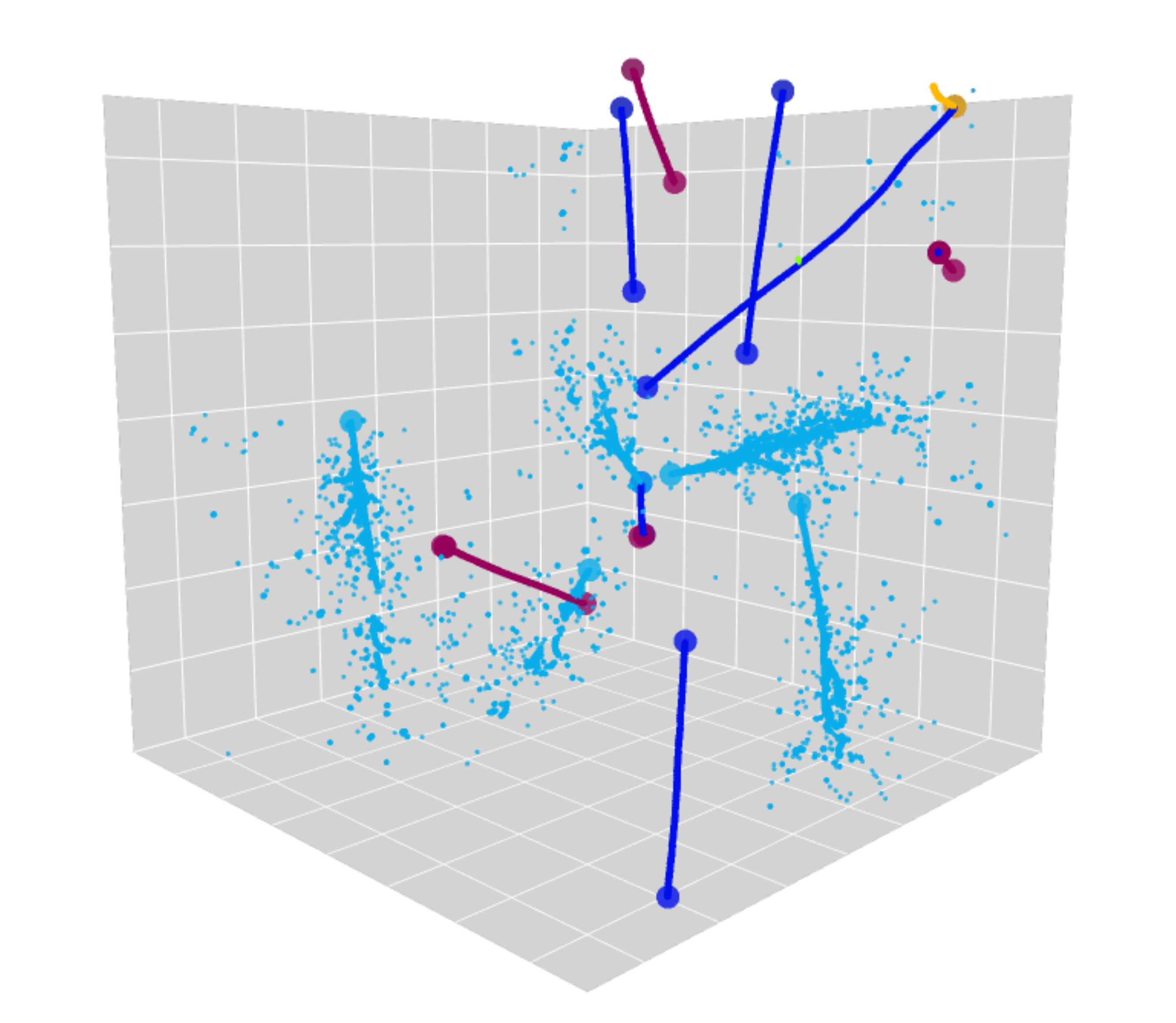}
        \caption{Simulated LArTPC event from our dataset. The left picture (network input) shows energy deposits from charged particle trajectories, whose color corresponds to an energy scale. In the right image (labels), each voxel is assigned one of five colors: heavily ionizing particles (HIP) in purple, minimum ionizing particles (MIP) in dark blue, electromagnetic showers in light blue, delta rays in green and Michel electrons in orange. }
    \label{fig:lartpc}
\end{figure*}

\subsection{Loss definitions}
Among all voxels $\vec{x_i}$ we define true voxels $A$ as voxels within a certain distance threshold $d_{positive}$ from the true points $\vec{q_j}$, and all other voxels as negative:
\[
A = \{ 1 \leq i \leq N \quad | \quad \exists j \quad\lVert  \vec{x_i} - \vec{q_j} \rVert < d_{\text{true}} \},
\]
where $N$ is the number of voxels in the input data tensor at a certain PPN layer.

We define several losses. First, for all input voxels, we compute a cross-entropy loss for positive/negative classification task at each PPN-$i$ layer and then average over all voxels. For $i = 1,2,3$, if $\vec{y}_k \in \{ 0; 1 \}$ indicates whether the voxel is positive or negative in the labels and $p_k$ is the predicted softmax score for this voxel to be positive,
    \[
        \mathcal{L}_{detection}^i = -\frac{1}{N_i} \sum_{k=1}^{N_i} y_k \log (p_k) + (1-y_k) \log (1-p_k).
    \]
    
Secondly, only on true voxels, we define a linear distance loss on the predicted positions. We consider the distance to the closest ground truth point $\vec{q}$. The raw predictions $\vec{p}$ of the network are actually shifts with respect to the center of the subject voxels ($0.5 + \vec{x}$):
    \[
        \mathcal{L}_{distance} = \frac{1}{N_3} \sum_{i = 1}^{N_3} \mathbbm{1}_{A} \min_{j} \lVert \vec{p}_i + 0.5 + \vec{x}_i - \vec{q}_j \rVert 
    \]
    
Thirdly, only on true voxels, we compute a cross-entropy loss for a point type prediction. The predicted point type is compared with the semantic type of the closest true point. If $N_c$ is the total number of semantic types for a point, $\vec{y}$ is a one-hot encoded vector indicating to which type the point belongs, and $p_c$ is the predicted probability that the point belongs to a semantic type $c$, then
    \[
        \mathcal{L}_{type} = - \frac{1}{N_3} \sum_{i=1}^{N_3} \mathbbm{1}_{A} \sum_{j=1}^{N_c} y_c \log (p_c).
    \]

Finally, the sum of all losses is minimized:
\[
\mathcal{L} = \mathcal{L}_{type} + \mathcal{L}_{distance} + \sum_{i=1,2,3} \mathcal{L}_{detection, i}
\]

\subsection{Post-processing}
The architecture that we proposed so far will yield a prediction of a position, detection score, and semantic type score for each voxel that has been selected in the last layer at the original image resolution. 
The number of such positive voxels whose predictions are considered, is related to the attention mask predicted by PPN2 and the spatial size ratio between PPN2 layer and the original image size. This will dictate for each voxel predicted as positive at PPN2 level how many voxels are selected at the last layer.
Hence we might have many proposals whose positions are clustered near a true point, with different scores and type predictions. We need a strategy to combine overlapping proposals to deduce the candidate of distinct 3D points, and we want this strategy to value both accurate positions and type predictions. In this paper, we adopt the following simple post-processing scheme.
\begin{enumerate}
\item Thresholding on the detection scores, for example we require a score value of 0.5 or higher to be considered positive. 
\item We then run the DBSCAN~\cite{DBScan} clustering algorithm on the positive point positions.  The hyper-parameters of DBSCAN are set to $\epsilon=$1.99 in voxel unit for the maximum Cartesian distance $\epsilon$ between two points to be clustered together, and \texttt{min\_samples}$=1$. $\epsilon$ must be small enough to avoid merging together predicted endpoints of short tracks.
\item Pooling operation on the points that belong to the same cluster in order to deduce a single score, type predictions, and 3D position. We use average-pooling for the 3D coordinate locations, and maximum-pooling for the scores including the positiveness prediction and individual semantic type. 
\item Finally, we enforce that a point detected by PPN as a type among $c_i$ (set of types, with type score $>0.5$ for each type $c_i$) needs to be within 2~voxels of a voxel predicted by U-ResNet to have one of the $c_i$ types.
\end{enumerate}

\section{Experiments\label{sec:experiments}}

\subsection{Dataset}
We use 3D LArTPC particle images from the PILArNet dataset~\cite{opendata}, an open dataset made available by the DeepLearnPhysics collaboration\footnote{\href{https://dx.doi.org/10.17605/OSF.IO/VRUZP}{https://dx.doi.org/10.17605/OSF.IO/VRUZP}}. We use the largest 3D image in the dataset, a cubic volume with each side 768 voxels (453 million voxels) at  3~mm/voxel spatial resolution. Figure~\ref{fig:lartpc} shows an example image from this dataset. The dataset contains 80,000 images for the training set and 20,000 images for the test set where each image contains several particles traversing the LAr volume. The PILArNet provides five types for the voxel-level semantic category. These include heavily ionizing particles (HIP, e.g. protons), minimum ionizing particles (MIP, e.g. muons and pions), electromagnetic (EM) showers, delta rays, and Michel electrons from muon decays. Further, the dataset provides particle-level metadata including endpoints of HIP and MIP particles as well as the initial point of other particle types including EM showers, delta rays and Michel electrons. These 3D points are provided with a floating-point precision in the unit of voxels, and used as true points for training PPN.  More details can be found in the PILArNet reference~\cite{opendata}.


\begin{figure}[t]
    \centering
    \includegraphics[width=0.98\linewidth]{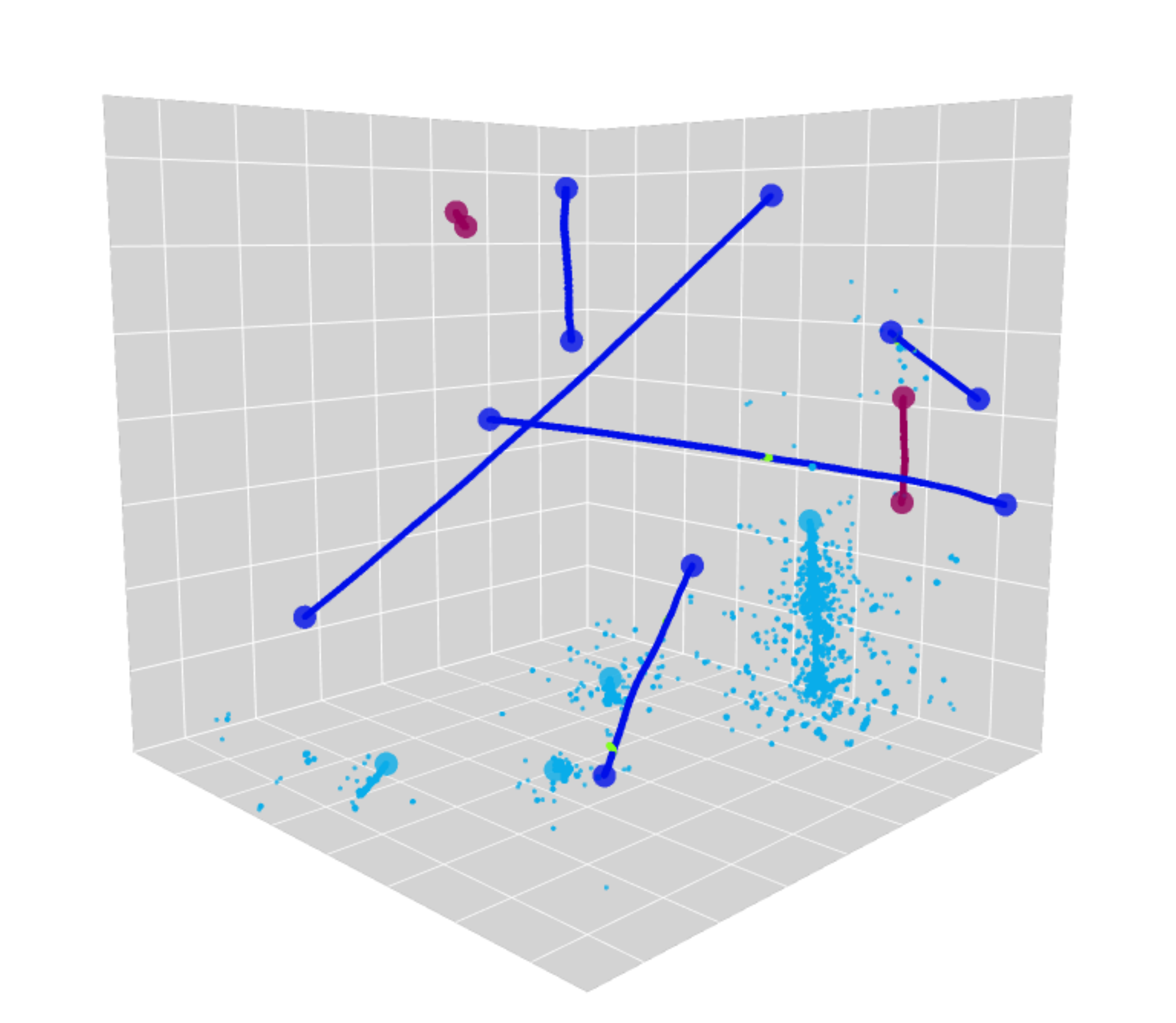}
    \caption{Example of predictions by UResNet+PPN. The voxels color corresponds to their semantic class as predicted by UResNet. The dots are proposed by PPN, and the dots color represents the point type predicted by PPN.}
    \label{fig:prediction}
\end{figure}

\subsection{Training details}
We drop the point labels for particles with less than 10~MeV in total energy deposit or a total voxel count of less than 7~voxels, which corresponds to a trajectory of a few voxels in length as a typical trajectory width is a few voxels or more.
The PPN1 and PPN2 layers have a spatial size of 24~voxels and 96~voxels respectively.
During training, we add true voxels to the attention masks generated by the PPN1 and PPN2 layers so that the subsequent layers, namely PPN2 and PPN3, can be trained with some mixtures of true and false voxels. This allows all PPN layers to train simultaneously from the beginning.


The batch size is 64  and we used an \textsc{Adam} optimizer with learning rate 0.001 to train the network.
Training the U-ResNet alone first for 20k iterations, then U-ResNet+PPN for another 20k iterations, took 184 hours on a Nvidia V-100 GPU for the total of 32~epochs.
The whole network (i.e. U-ResNet+PPN) can be trained from scratch without having to separate into two stages, in which case 40k iterations took 231 hours. Unless stipulated otherwise, the default configuration for the rest of this paper is the two-stage training.

\section{Results \label{sec:results}}

\begin{figure}[t]
    \centering
    \includegraphics[width=0.98\linewidth]{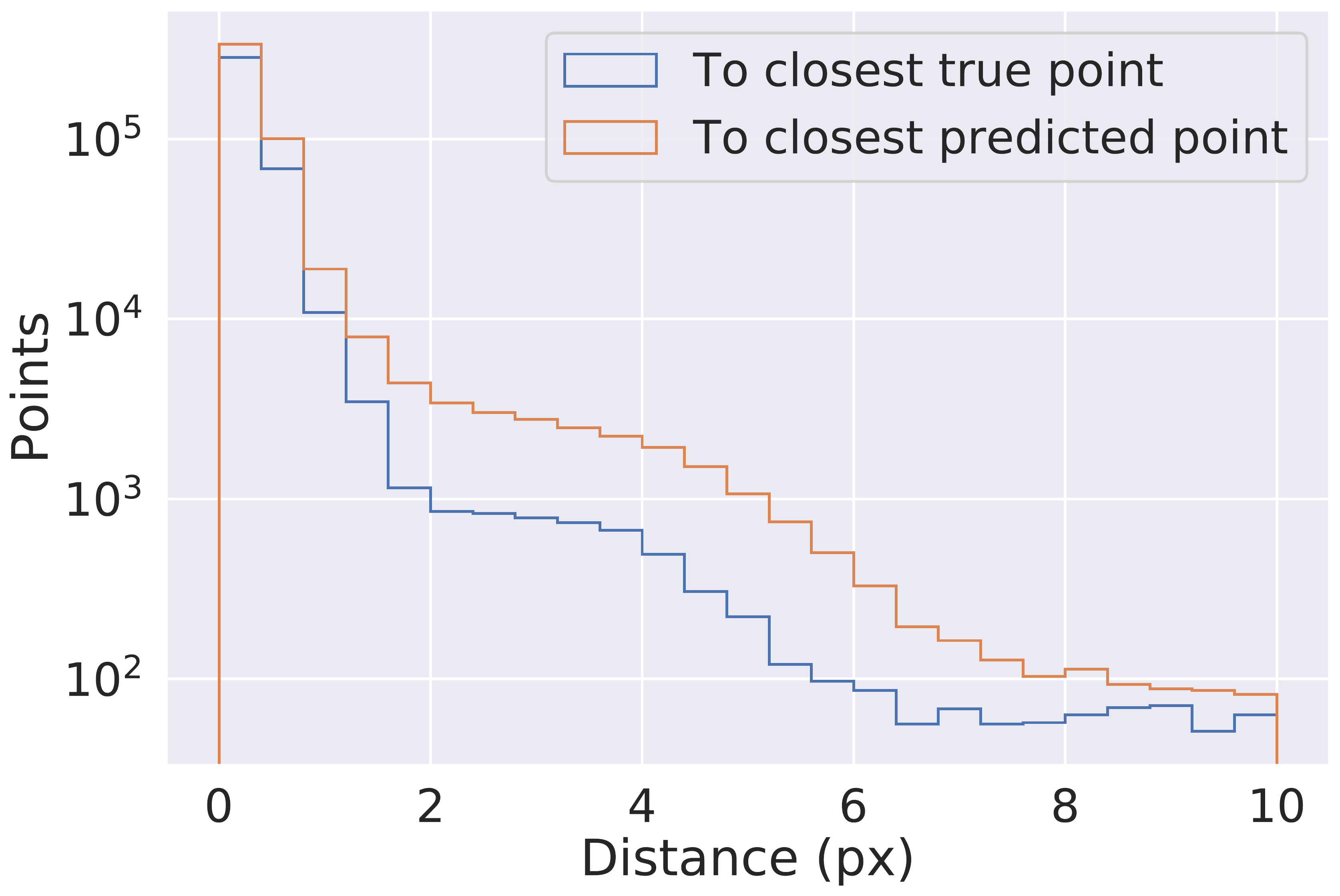}
    \caption{Distance from true points to the closest predicted points, and from predicted points to the closest true points. Both the true and predicted points of delta ray type are excluded in this plot. 97.8~\% of the points are contained in the x-axis range for both histograms.}
    \label{fig:distance}
\end{figure}

\begin{figure}[t]
    \centering
    \includegraphics[width=0.98\linewidth]{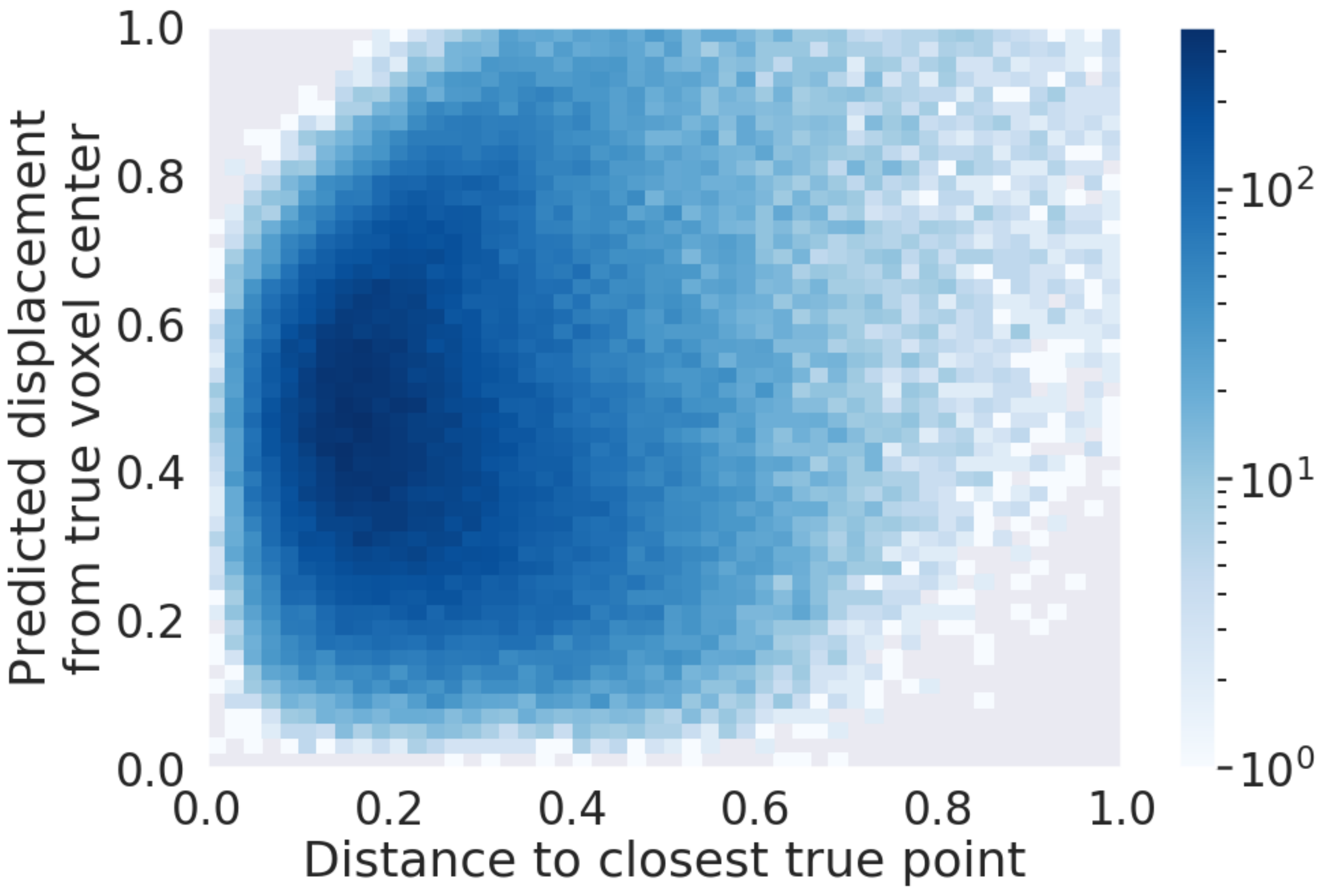}
    \caption{Correlation between the distance from a predicted point to the closest true point, and the distance from a predicted point to the voxel center of the corresponding true point. We selected predicted points that are within 3~voxels of a true point. About 3.6~\% of these predicted points are associated with true points that PILArNet locates exactly at the center of a voxel. They are not pictured here.}
    \label{fig:pixelresolution}
\end{figure}


\begin{figure}[t]
    \centering
    \includegraphics[width=0.98\linewidth]{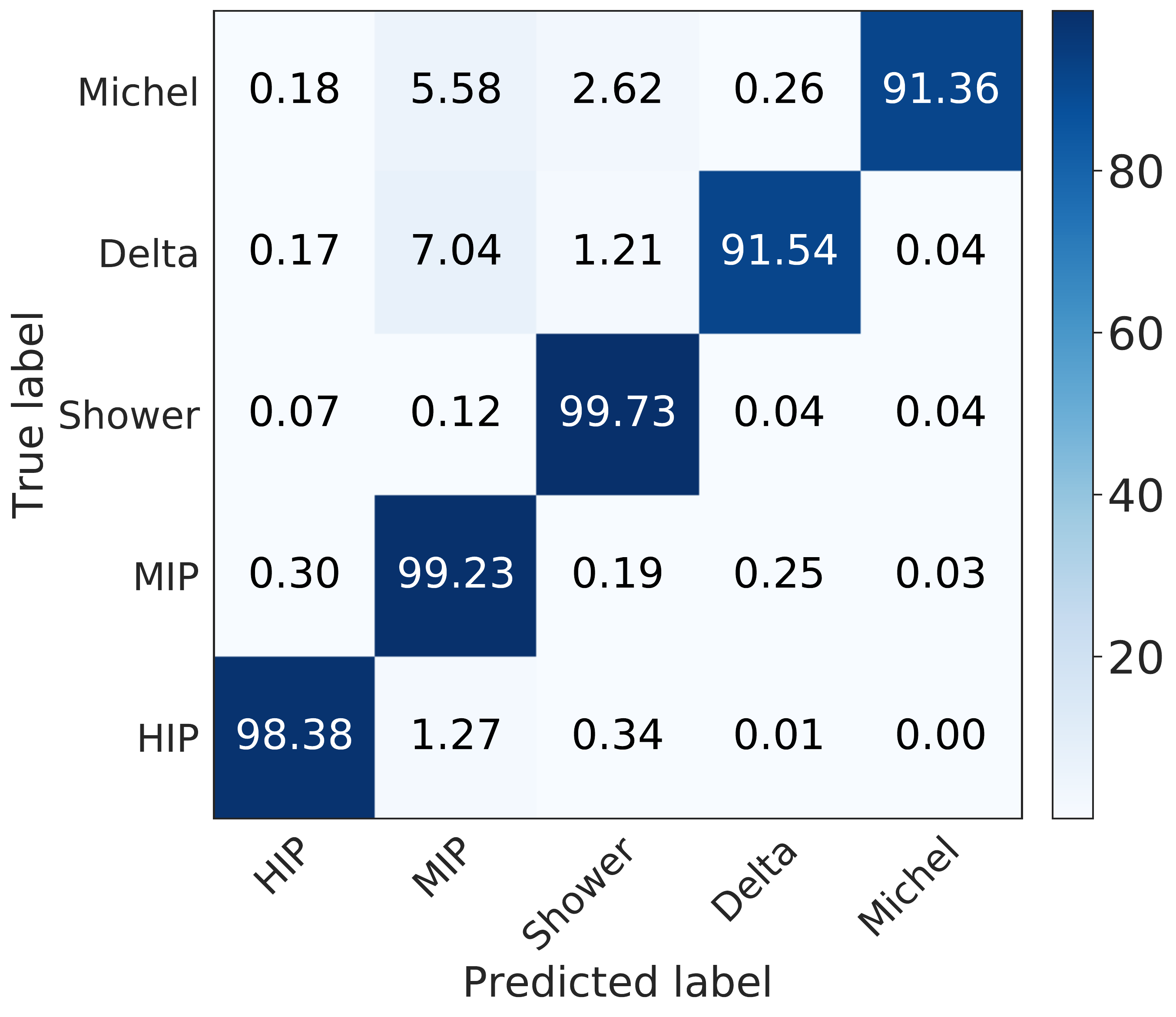}
    \caption{Confusion matrix for U-ResNet. Each cell contains the fraction of voxels in percent belonging to a certain true semantic type on the vertical axis that have been predicted as the semantic type shown on the horizontal axis. Each row sums to 100~\%.}
    \label{fig:confusion_matrix}
\end{figure}

\subsection{Position Precision}
Figure~\ref{fig:prediction} is a visual example of predictions made by UResNet+PPN.
Figure~\ref{fig:distance} shows the distribution of distance from a true point to the closest predicted point. For all true points, 95.1~\% and 97.8~\% of them are within the voxel distance of 3 and 10 from a predicted point. The figure also shows the distribution of distances from a predicted point to the closest true point, and we find that our algorithm successfully predicts 96.8~\% and 97.8~\% of 3D points within the voxel distance of 3 and 10 from the true points respectively. If we look at semantic type-wise results, we find that the fraction of true points which are more than 3~voxels away from any predicted point is 7~\%, 2.1~\%, 8.2~\% and 1.6~\% for the HIP, MIP, EM shower and Michel electron types respectively. For this analysis and Figure~\ref{fig:distance}, we excluded delta ray type true and predicted points since the true point coordinates for delta rays provided in PILArNet are less precise than those of MIP, HIP, EM shower and Michel electron types.  This is likely because the initial points of delta rays often overlap with a muon trajectory, which typically has a width of a few voxels or more. We considered the predicted points to be delta ray type if the point has the delta ray type score of 0.5 or higher.

 For those predicted points within 3~voxels of the closest true point, the median distance between the positions of a predicted point and the closest true point is measured to be 0.25~voxels. If PPN is only sensitive to identify a true voxel in which a true point is present, and if it is not capable of regressing the position at the sub-pixel level, we expect this resolution to be 0.66, which is the median distance between two random point positions in a voxel. We note that 17~\% of the true points provided by the PILArNet are located exactly at the center of a voxel, which is typically observed as a result of an endpoint approximation within the recorded volume when a particle is exiting or entering the volume. For other true points whose position is not fixed exactly at the voxel center, the correlation of the distance between a predicted point to the closest true point and the displacement of that predicted point from the true voxel center is shown in Figure~\ref{fig:pixelresolution}. When PPN predicts a point within the correct true voxel, geometrically the distance from the voxel center to the predicted point must be between 0 and 0.866.  The two distances show almost no correlation in this range, which shows that PPN position resolution is uniform and independent of a true point location within the true voxel. This demonstrates that our algorithm achieved a sub-voxel level precision for this reconstruction task.

\subsection{Type Prediction Accuracy}

\begin{figure}[t]
    \centering
    \includegraphics[width=0.98\linewidth]{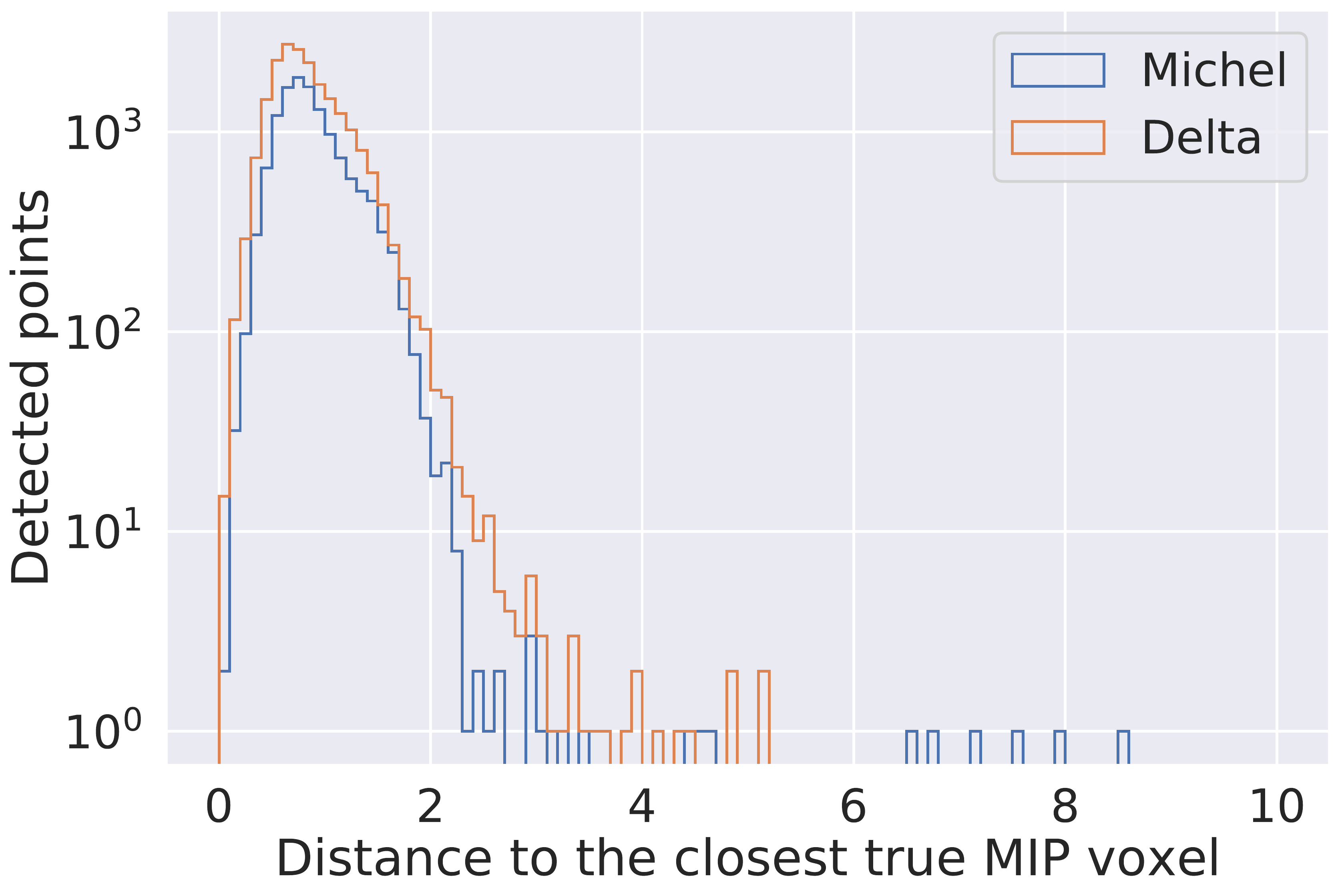}
    \caption{Looking at the distance between predicted points of Michel electron or delta ray type (i.e. with a corresponding semantic type score $> 0.5$) and the closest voxel with the true semantic type of MIP.}
    \label{fig:michel_delta}
\end{figure}

\begin{figure}[t]
    \centering
    \includegraphics[width=0.98\linewidth]{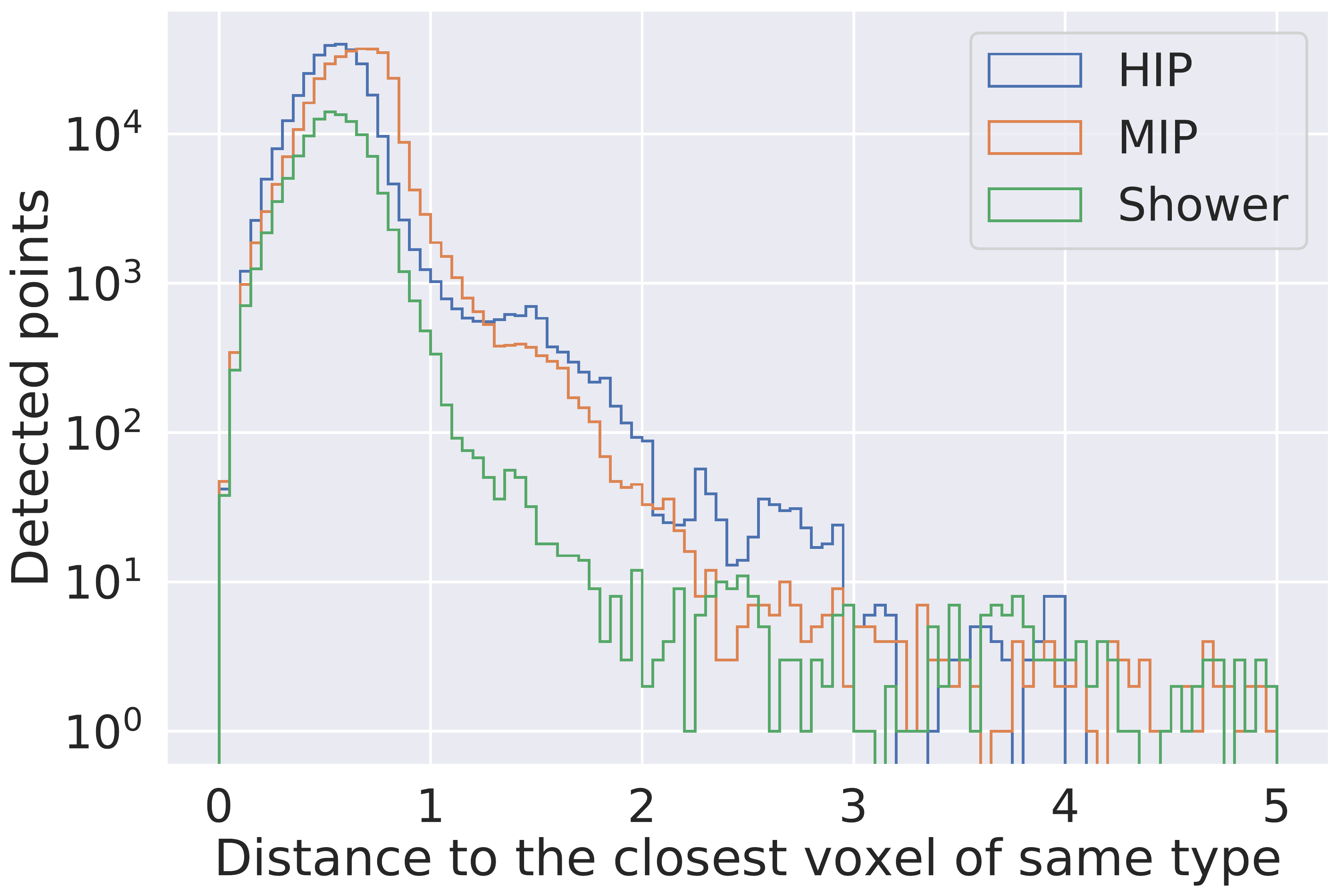}
    \caption{This histogram shows the distance from a predicted point before post-processing with a type HIP, MIP, or EM shower (i.e. type score $> 0.5$) to the closest voxel of the same type as predicted by U-ResNet. 1024 events are used in this histogram.}
    \label{fig:closest_pred_pix}
\end{figure}

Before evaluating the point type prediction performance, it is useful to remind ourselves of the performance of the U-ResNet. In this particular training, the confusion matrix that we obtain is shown in Figure~\ref{fig:confusion_matrix}.
We can then look at the distance from predicted points to true voxels of a certain semantic type. For example we expect predicted points with high Michel or Delta type score to be close to MIP voxels in the labels, and Figure \ref{fig:michel_delta} confirms this.
Figure~\ref{fig:closest_pred_pix} shows that imposing the maximal distance threshold of 2~voxels between a predicted point with a high type score for a set of types and a voxel whose predicted type matches one of them is reasonable.

We define purity and efficiency metrics for the point type prediction as follows: for a given predicted point, we consider all semantic types for which it has a score $> 0.5$ and we refer to them as predicted types. We count a predicted type as matched if there exist a true point of the same type within 5~voxels. We note that one point may be associated with multiple types, and one  point may contribute as many times as its associated type counts under this scheme.
The fraction of matched predicted types is our purity metric. Similarly, for a given true label type, we say that it is matched if there is a predicted point within 5~voxels which has a score $> 0.5$ for the same semantic type. The fraction of matched true types is our efficiency metric. Under these definitions, we find a purity of 96.3~\% and an efficiency of 89.2~\%.
The Table~\ref{tab:type_metrics} breaks down these purity and efficiency metrics for each semantic class. The purity is significantly higher than the efficiency, which indicates that while currently predicted types are highly accurate, there is a space for PPN to improve to predict all possible types associated with a predicted point. This may be related to the architecture where PPN is coupled to U-ResNet which ultimately predicts a single type per voxel.

\begin{table}[t]
    \centering
    \begin{tabular}{ccccc}
        \hline \hline
        Class & HIP & MIP & Shower & Michel \\
        \hline
        Purity & 97~\% & 98~\% & 92.2~\% & 99.3~\% \\
        Efficiency & 85.3~\% & 93.4~\% & 90.7~\% & 89.7~\% \\
         \hline \hline
    \end{tabular}
    \caption{Purity and efficiency of the point type predictions for different semantic classes. Purity for a class X is the fraction of predicted points with a type score $>0.5$ for this class X, and within 5~voxels of a true point whose type matches one of the predicted point types (i.e. type score $>0.5$). Efficiency for a class X is the fraction of true points of type X within 5~voxels of a predicted point with type score $>0.5$ for this class X.}
    \label{tab:type_metrics}
\end{table}


\subsection{Mistakes analysis}
About 2.2~\% of predicted points, excluding points predicted as delta ray type, are more than 10~voxels away from any true point. Let us call them far mistakes. Among them, 25.8~\% have a high ($>0.5$) type score of being HIP, 21.9~\% for MIP and 53.8~\% for shower. We have visually scanned event displays of these mistakes and report their nature in this section. In summary, we found that a large fraction of far mistakes were due to issues with true points or legitimate mistakes where authors cannot visually distinguish from correct predictions.



\subsubsection{Fragmented EM Showers}
An EM shower is initiated by an EM particle including an electron, a positron, or a gamma ray, and develops a cascade of them through radiations of gamma rays. In physics analysis, typically a whole cascade is conveniently treated as one EM shower instance instead of identifying dozens to hundreds of secondaries. This is shared in the PILArNet dataset where EM shower information, such as the initial point, is provided for the whole cascade. In LArTPC, however, given the average radiation length of 14~cm~\cite{radlength}, which corresponds to 47 voxels, we expect that some radiated gamma rays in the cascade may be separated by significant gaps. This results in cases where a single EM shower may appear indistinguishable from two or more separate, overlapping EM showers.  

\begin{figure}
    \centering
    \includegraphics[width=0.9\linewidth]{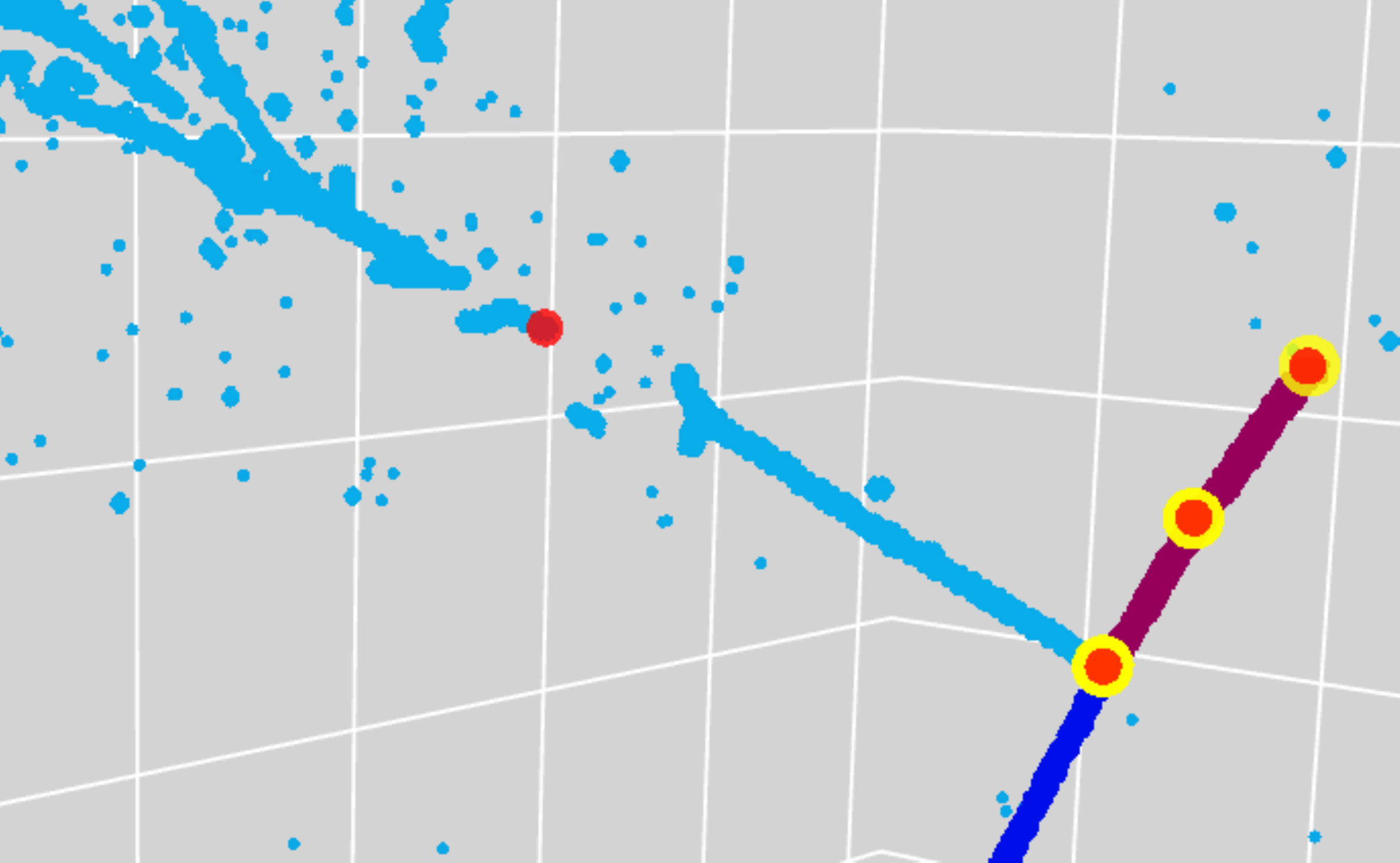}
    \caption{This picture shows the semantic and point predictions of UResNet and PPN. Predicted EM shower voxels are in cyan, MIP in dark blue and HIP in purple. The yellow points are true points. The red points are predicted by PPN. The shower fragment on the left belongs to the EM shower coming from the right, and PILArNet only provides a single initial point for the whole shower (the yellow dot at the shower start on the right, where PPN correctly predicted another EM shower point).}
    \label{fig:mistakes_shower}
\end{figure}

In those cases, PPN may place multiple initial points within a single shower, as shown for example in Figure~\ref{fig:mistakes_shower}. While this may visually appear reasonable, they can be the cause of far mistakes as the PILArNet provides only one initial point for the whole shower. Among the far mistakes, more than half (53.2~\%) have a high ($>0.5$) type score for being EM shower and are within 2~voxels distance from voxels of true shower type in semantic segmentation labels. 

\subsubsection{Mistakes due to tracks (HIP/MIP)}
\begin{figure}[t]
    \centering
    \includegraphics[angle=90,width=0.7\linewidth]{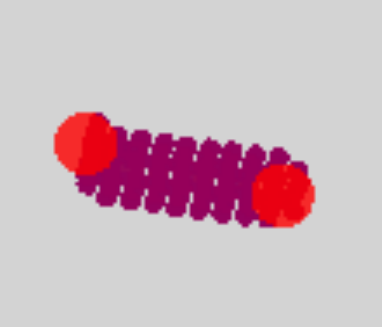}
    \caption{Example of a short trajectory (purple voxels) that is lacking true points due to small total energy deposit. The red points are predicted by PPN.}
    \label{fig:small_cluster}
\end{figure}

\begin{figure}[t]
    \centering
    \includegraphics[angle=90,width=0.7\linewidth]{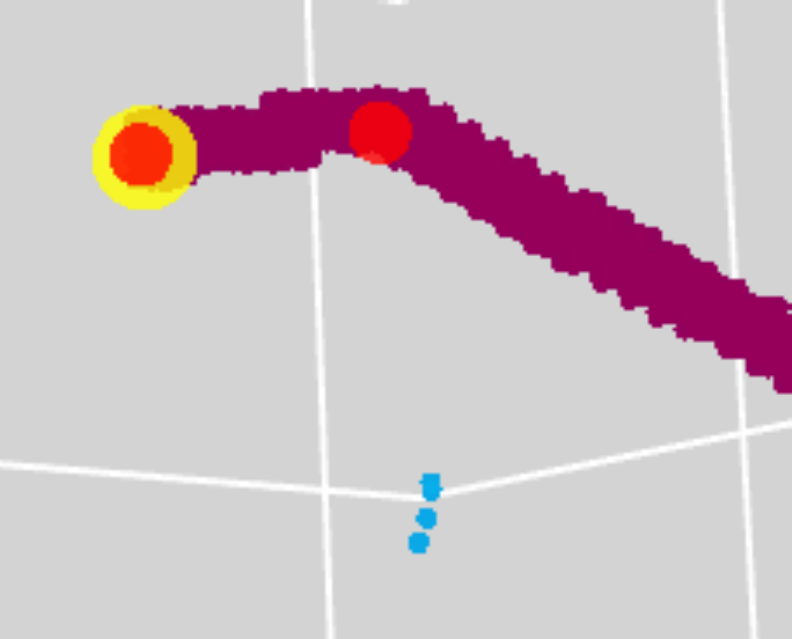}
    \caption{Example of kink along a HIP trajectory (purple voxels) causing a legitimate mistake by PPN. The yellow points are true points. The red points are predicted by PPN.}
    \label{fig:kinks}
\end{figure}

49.4~\% of the far mistakes have a HIP or MIP point type score $>0.5$.
We randomly sampled 20 cases with one far mistake with high HIP score. 
15 of them (75~\%) were due to very small HIP trajectories for which PPN made good predictions but true points were missing. This is caused by the fact that these trajectories fall below the threshold that we impose to define true points ($>7$ voxels, $>10$~MeV total energy deposit), leading to missing true points as shown in Figure~\ref{fig:small_cluster}. 2 out of the remaining 5 cases were found to be ``legitimate mistakes'' due to a kink in a trajectory as shown in Figure~\ref{fig:kinks}.  The last 3 cases were genuine mistakes (e.g. a point predicted in the middle of a trajectory without any obvious kink).


On the other hand we also sampled 20 events with far mistakes with high MIP score. One case was due to a short trajectory missing true points, and one was a rare case where PPN made an extra, faulty prediction at the crossing point of two MIP trajectories that accidentally overlapped in the 3D space. The majority (12 cases) were legitimate mistakes due to a kink in a trajectory. The rest (6 cases) were genuinely bad mistakes. 

\subsubsection{Trajectories affected by the boundaries}
10.1~\% of far mistakes are within 5~voxels of an image boundary, indicating they may come from a particle trajectory crossing the image volume boundary. MIP trajectories are more likely to cross a volume boundary due to their length. Hence they are more affected by boundaries. Among the far mistakes that are more than 5~voxels away from the boundary, only 18~\% have a high MIP type score. This fraction increases to 54~\% in the region within 5~voxels from the boundary while negligible statistical change was observed for predicted points of other types.

\begin{figure}[t]
    \centering
    \includegraphics[angle=90,width=0.7\linewidth]{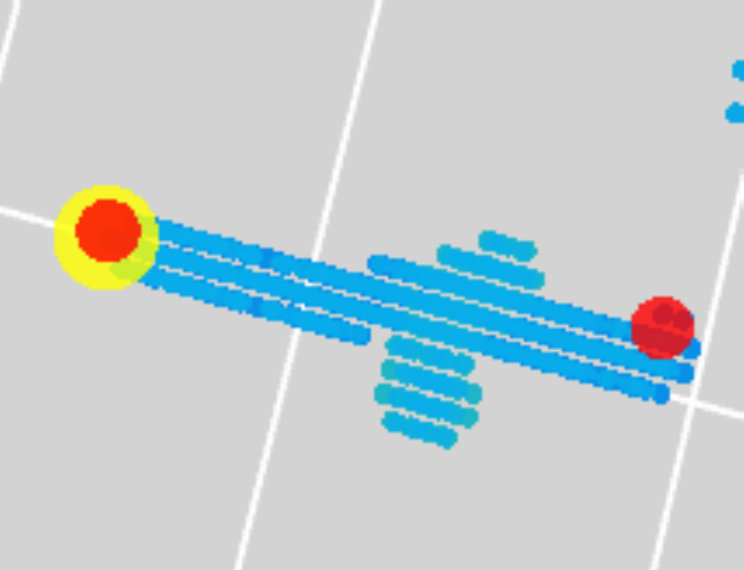}
    \caption{Example of shower trajectory exiting the volume. All the cyan (true shower) points were mistakenly classified as MIP by UResNet. The yellow points are true points. The red points are PPN predictions with high MIP score.}
    \label{fig:exiting}
\end{figure}

\begin{figure}[t]
    \centering
    \includegraphics[angle=90,width=1.0\linewidth]{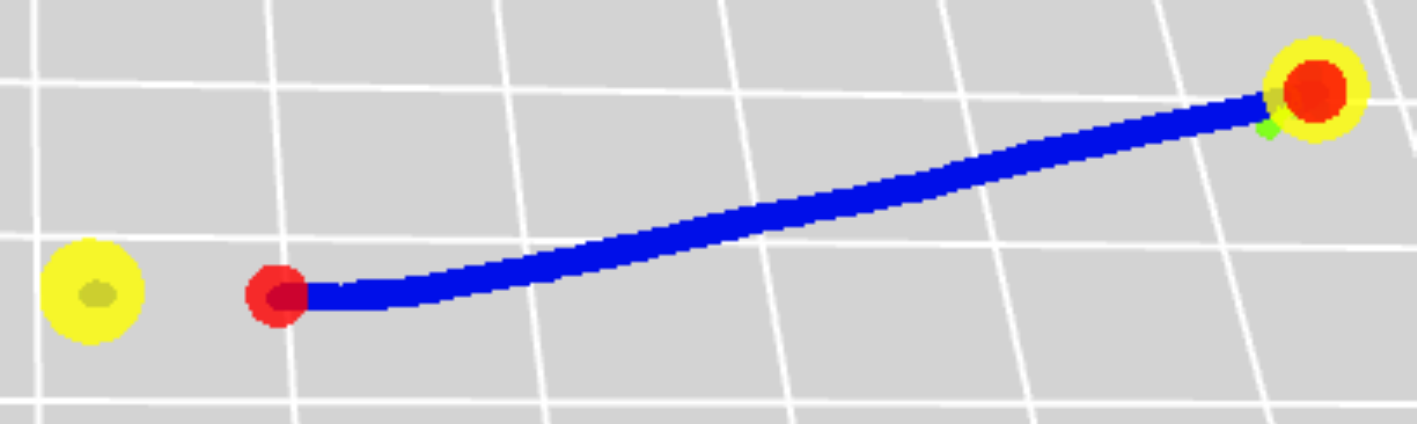}
    \caption{Example of MIP trajectory exiting and briefly re-entering the volume. The yellow points are label points. The red points are PPN prediction.}
    \label{fig:reentering}
\end{figure}

\begin{figure}[t]
    \centering
    \includegraphics[width=1.0\linewidth]{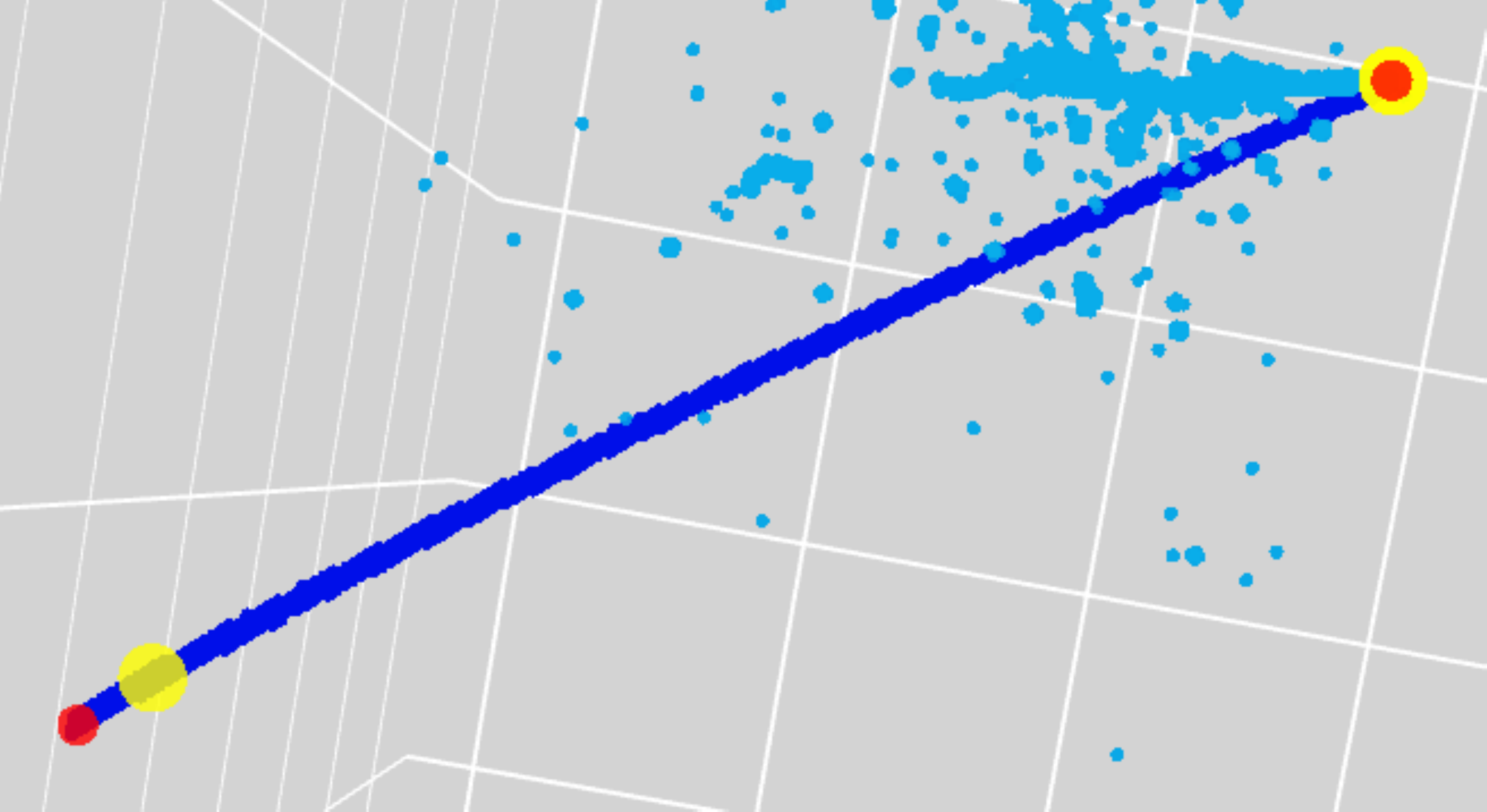}
    \caption{An example issue of a true point that appears to be mistakenly shifted. The yellow points are true label points. The red points are PPN prediction.}
    \label{fig:label_shifted}
\end{figure}

We have visually scanned randomly selected 10 far mistakes of a high MIP type score in this region next to the boundary. One of them was a legitimate mistake due to a kink in a trajectory, similar to the dominant case of MIP far mistakes found and described previously. The rest (9) of the MIP mistakes near the boundary were all due to issues related to true points. These issues include: exiting shower trajectory which gets classified as MIP by UResNet (Figure~\ref{fig:exiting}), and results in a too short trajectory and loss of true points as previously described for HIP cases, a MIP trajectory that exited and re-entered the image volume (Figure~\ref{fig:reentering}),  for which the true points provided in PILArNet appear unreasonable, and also what appears as a genuine mistake of true point location on the boundary provided by PILArNet dataset (Figure~\ref{fig:label_shifted}). We conclude therefore that the majority of far mistakes made by PPN are due to either issues related to true points or legitimate mistakes that visually appear reasonable.


\subsection{Others}
\begin{table}[t]
    \centering
    \begin{tabular}{ccccc}
    \hline \hline
    & \multicolumn{2}{c}{Duration (s)} & \multicolumn{2}{c}{Memory (GB)} \\
    \hline
    & Train & Test & Train & Test \\
    \hline
    UResNet only & 14.3 & 4.9 & 9.9 & 2.2\\
    UResNet + PPN & 20.5 & 7.6 & 10.8 & 2.2\\
    \hline \hline
    \end{tabular}
    \caption{Resources usage in time and memory of UResNet and UResNet+PPN architectures, on Nvidia V-100 GPUs.}
    \label{tab:resources}
\end{table}

We also compared the PPN performance in two training scenarios: a single-stage training, where we start training from scratch both U-ResNet and PPN at the same time for 40k iterations, and a two-stage training where we train U-ResNet for 20k iterations first, before adding the PPN layer and continue training of U-ResNet+PPN for 20k more iterations. Everything else is identical between the two schemes. The fraction of true points that are within 10~voxels of a predicted point is 98.2~\% and 97.8~\% respectively. The fraction of predicted points that are within 10~voxels of a true point is 97.8~\% in both cases.
The fraction of true points which are more than 3~voxels away from any predicted point is 5.2~\% and 5.4~\% respectively.
There is no significant difference between the two training schemes, which confirms that the PPN learning is conditioned by the UResNet performance.



Table~\ref{tab:resources} shows that PPN layers have a very little impact on the memory usage (about 1GB at train time, negligible at inference time). However they are responsible for about 30~\% of the total computation time, if compared with the UResNet-only resources usage.

\subsection{Track clustering}
\begin{figure*}[t]
    \centering
    \includegraphics[width=0.98\textwidth]{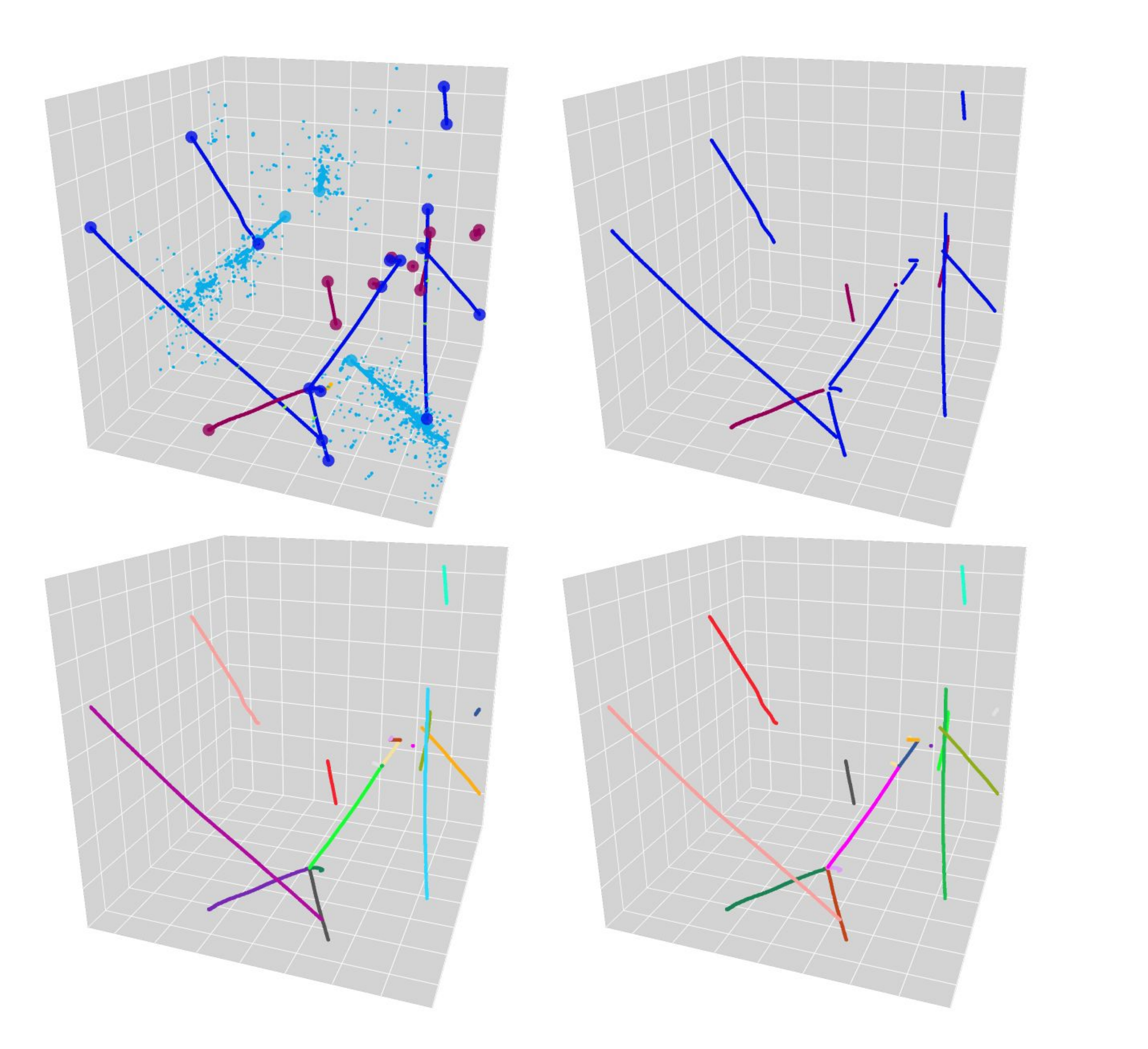}
    \caption{Track clustering. Top left: UResNet + PPN predictions. Top right: selecting UResNet track predictions only and removing voxels around PPN predictions. A radius of 10~voxels is used to make the gaps visible in this figure, but the clustering algorithms use a radius of 7~voxels. Bottom left: true track particle clusters. Bottom right: predicted track particle clusters.}
    \label{fig:trackclustering}
\end{figure*}
Lastly we report a simple application of U-ResNet and PPN for clustering voxels to identify individual track-like particles.
This clustering task belongs to the next important step in the LArTPC data reconstruction pipeline. Using the output of UResNet and PPN, a very simple clustering algorithm can be designed: first, for each predicted semantic type, run a density-based clustering algorithm such as DBSCAN \cite{DBScan} on the voxels predicted to belong to track-like particles (i.e. HIP and MIP types). We use here the parameters of $\epsilon = 4$ and \texttt{min\_samples}$ = 7$ for DBSCAN. This will cluster together particle trajectories that are spatially adjacent, such as tracks coming out of the same interaction vertex. To mitigate this issue we can use the points predicted by PPN to ``break" the predicted clusters: for each predicted cluster from the first step and associated closest predicted points, we mask a sphere of 7~voxels around each predicted point, run DBSCAN again to reconstruct the main trunk of individual track-like particles, and assign the remaining voxels in the masked regions to the closest track-like cluster to complete individual trajectories. Figure~\ref{fig:trackclustering} illustrates this simple algorithm.

We define metrics of purity and efficiency per cluster, as fraction of the predicted cluster voxels overlapping with the true cluster and fraction of the true cluster voxels overlapping with the predicted cluster respectively.  We also look at the Adjusted Rand Index (ARI) metric~\cite{ARI} per true semantic type (HIP and MIP), averaged over events, to get a sense for the overall clustering performance. We find for efficiency/purity/ARI metrics the values of 0.96/0.93/0.91 for track-like clusters.

\section{Conclusion}
We have introduced the Point Proposal Network. Building on the previous development of U-ResNet~\cite{domine2019scalable}, we showed that PPN is capable detecting the endpoints of track-like particles as well as the initial point of shower-like particles. PPN successfully predict 96.8~\% and 97.8~\% of 3D points within the voxel distance of 3 and 10 from the true points respectively. For the predicted points and true points that reside within 3~voxels within each other, PPN achieves the sub-voxel level precision with a median distance of 0.25~voxels. PPN is also the first benchmark algorithm for PILArNet for reconstructing particle positions. Using the output of U-ResNet and PPN, we demonstrated a simple set of algorithms to cluster 3D voxels into individual track-like particles. We reported that our algorithms achieved a voxel clustering efficiency/purity/ARI of 0.96/0.93/0.91. U-ResNet and PPN are part of a scalable, deep-learning based data reconstruction chain for LArTPC detectors.

\section{Acknowledgement}
This work is supported by the U.S. Department of Energy, Office of Science, Office of High Energy Physics, and Early Career Research Program under Contract DE-AC02-76SF00515.

\bibliography{references}
\end{document}